\documentclass[a4paper,oneside,final,notitlepage,onecolumn,12pt]{article}
\usepackage{amsfonts}
\usepackage{epsf}
\usepackage{graphicx}% Include figure files
\usepackage{amssymb,eso-pic}
\usepackage{latexsym}
\usepackage{tabularx}
\usepackage{amsxtra}
\usepackage{hyperref}
\usepackage{t1enc}
\usepackage{ifpdf,epsfig,array,amsmath,amssymb,psfrag}

\usepackage[normalem]{ulem}
\usepackage{color}
\definecolor{blue}{rgb}{0,0,1}
\definecolor{red}{rgb}{1,0,0}


\DeclareFontFamily{OT1}{rsfs}{} \DeclareFontShape{OT1}{rsfs}{m}{n}{
<-7> rsfs5 <7-10> rsfs7 <10-> rsfs10}{}
\DeclareMathAlphabet{\mycal}{OT1}{rsfs}{m}{n}

\def\scri{{\mycal I}}%skraj

\def\sc{{\hskip 3.5pt {{}^{{}^{{}_{{}_{\bowtie}}}}} \kern -8.pt{}}}  
\def\SC{{\hskip 3.5pt {{}^{{}^{{}^{{}_{{}_{\bowtie}}}}}} \kern -10.5pt{}}}

\begin{document}

%%%%%%%%%%%%%%%%%%%%%%%%%%%%%%%%%%%%%%%%%%%%%%%%%%%%%%%%%%%%%%%%%%%%%%%%%%%%%%
\newtheorem{theorem}{Theorem}[section]
\newtheorem{lemma}{Lemma}[section]
\newtheorem{proposition}{Proposition}[section]
\newtheorem{corollary}{Corollary}[section]
\newtheorem{conjecture}{Conjecture}[section]
\newtheorem{example}{Example}[section]
\newtheorem{definition}{Definition}[section]
\newtheorem{remark}{Remark}[section]
\newtheorem{exercise}{Exercise}[section]
\newtheorem{axiom}{Axiom}[section]
%%%%%%%%%%%%%%%%%%%%%%%%%%%%%%%%%%%%%%%%%%%%%%%%%%%%%%%%%%%%%%%%%%%%%%%%%%%%%%
\renewcommand{\theequation}{\thesection.\arabic{equation}} 
% A fenti parancs atdefinialja az egyenleteket szamozo parancsot
%%%%%%%%%%%%%%%%%%%%%%%%%%%%%%%%%%%%%%%%%%%%%%%%%%%%%%%%%%%%%%%%%%%%%%%%%%%%%%

\author{Andor Frenkel\,%\thanks{% ~email: frenkel.andor@wigner.mta.hu} 
\  and \ Istv\'an R\'{a}cz\,%\thanks{% ~email: racz.istvan@wigner.mta.hu}  
\\ %EndAName 
Wigner RCP, \\ H-1121
Budapest, Konkoly Thege Mikl\'os \'ut 29-33. \\Hungary
}
\title{On the use of projection operators \\ in electrodynamics} 
%\title{On the projection operator determining \\ the Coulomb gauge vector potential} 
%\title{Potentials for moving point charges \\ in Lorenz and Coulomb gauges} 
\maketitle

\begin{abstract}
In classical electrodynamics all the measurable quantities can be derived from the gauge invariant Faraday tensor $F_{\alpha\beta}$. Nevertheless, it is often advantageous to work with gauge dependent variables. In \cite{jackson2}, \cite{hnizdo} and \cite{wundt}, and in the present note too, the transformation of the vector potential in Lorenz gauge to that in Coulomb gauge is considered. This transformation can be done by applying a projection operator that extracts the transverse part of spatial vectors. In many circumstances the proper projection operator is replaced by a simplified transverse one. It is widely held that such a replacement does not affect the result in the radiation zone. In this paper the action of the proper and simplified transverse projections will be compared by making use of specific examples of a moving point charge. It will be demonstrated that whenever the interminable spatial motion of the source is unbounded with respect to the reference frame of the observer the replacement 
of the proper projection operator by the simplified transverse one yields, even in the radiation zone, an erroneous result with error which is of the same order as the proper Coulomb gauge vector potential itself.    
\end{abstract}

\vfill\eject

\section{Introduction}
\setcounter{equation}{0}

Consider a spatial vector field $\vec{V}$. Its transverse---i.e.~divergence free---part $\vec{V}_T$ can be determined by making use of the projection operator\,\footnote{It can be justified by applying the relation $\nabla^2_{\bf x}\,\frac{1}{|{\bf x}-{\bf x}'|}=-4\pi\,\delta({\bf x}-{\bf x}')$ that $P$, as given by (\ref{projT}), does indeed extract the transverse part of $\vec{V}$, i.e.\,$P[\vec{V}] = \vec{V}_T$ satisfies the relation $\vec{\nabla}\cdot \vec{V}_T=0$.} $P$ given as \cite{jackson} 
\begin{equation}\label{projT}
P[\vec{V}]=\vec{V}+ \frac{1}{4\pi}\,\vec{\nabla}\cdot\left[ \int\frac{\vec{\nabla}_{\vec{r}{\,}'}\cdot\vec{V}(\vec{r}{\,}') }{ |\vec{r}-\vec{r}{\,}'|}  \,d\vec{r}{\,}'\right] \,.
\end{equation}

In the radiation zone instead of this operator a simplified transverse one   
\begin{equation}\label{projN}
{\not\hskip-0.1cm P} [\vec{V}]   = \vec{V} -  (\vec{n}\cdot\vec{V})\,\vec{n} %\,,
\end{equation}
is applied, where the spatial unit vector $\vec{n}$ is pointing from the source to the point of observation. Notice that by construction ${\not\hskip-0.1cm P} [\vec{V}]$ is transverse to $\vec{n}$ but, in general, it is not divergence free. Our main concern in this paper is to demonstrate that such a replacement, as opposed to the ``folklore'', may lead to erroneous result. As a measure of the discrepancy of the action of the proper and the simplified transverse projection operators on $\vec{V}$ one may use $\Delta[\vec{V}]$ defined as 
\begin{equation}\label{projD}
\Delta[\vec{V}]=P[\vec{V}]-{\not\hskip-0.1cm P} [\vec{V}]   = (\vec{n}\cdot\vec{V})\,\vec{n}+\frac{1}{4\pi}\,\vec{\nabla}\cdot\left[ \int\frac{\vec{\nabla}_{\vec{r}{\,}'}\cdot\vec{V}(\vec{r}{\,}') }{ |\vec{r}-\vec{r}{\,}'|}\,  d\vec{r}{\,}'\right] \,.
\end{equation}

\medskip

In this paper it will be shown that the replacement of the proper projection operator $P$ by the simplified transverse one ${\not\hskip-0.1cm P}$ may be erroneous in determining the transverse (or Coulomb gauge) part of the vector potential of an electromagnetic filed in certain physically interesting situations. More concretely, explicit examples will be investigated, each with a point charge moving on a predetermined orbit (i.e.\,the back reaction will be left out from the present considerations), and it will be demonstrated that whenever the interminable spatial motion of the source is unbounded with respect to the reference frame of the observer the replacement of the proper projection operator by the simplified transverse one does indeed yield an erroneous result and the error is of the same order as the proper Coulomb gauge vector potential itself even if the comparison is made in the radiation zone.   

\medskip

Before proceeding it is important to recall that in the literature two types of asymptotic limits are applied. In the conventional approach the asymptotic limit is meant to be done by picking a $t=const$ time-slice and assuming that the distance of the observation point with respect to the origin tends to infinity. Notice that the retarded time inevitably tends to $-\infty$ in such a limiting process. Thereby, while the involved observation points are getting farther and farther from the observer they report us about earlier and earlier parts of the history of the source. 

In the other approach the retarded time $t_{ret}$ is kept constant and the observations are assumed to be done further and further away to the future along a null line originating at a distinguished event of the world line of the source determined by its location at the moment $t_{ret}$. This approach suits more to the investigation of radiative processes and it provides a meaningful determination of the distance dependence of the potential. The infinite limit along the chosen null line corresponds to an ideal endpoint of this line, representing a point---in the conformal setup introduced by Penrose \cite{penrose}---at future null infinity, $\scri^+$. In all of the considered particular cases we shall indicate the limiting behavior in this latter sense. 

\medskip

This paper is organized as follows: 
Section\,\ref{prel} is to select the class of electromagnetic systems to which our results apply and to recall some of the basic notions and variables we shall use. The justification of the key formulas is presented separately in the Appendix. The asymptotic behavior of the proper projection and its discrepancy from the simplified one for a point-like charge moving on predetermined orbits will be discussed in Section\,\ref{cases}.  In particular, the constant velocity and the oscillatory motions, along with their superposition will be investigated in some details. Section\,\ref{con} contains our final remarks. 

\section{Preliminaries}\label{prel}
\setcounter{equation}{0}

The electromagnetic field is represented by a $2$-form field $F_{\,\alpha\beta}$ satisfying the Maxwell equations 
\begin{equation}\label{m1}
\partial^\alpha F_{\,\alpha\beta}=-4\pi\,J_{\beta}\ \ \ {\rm and}\ \ \ \partial_{[\alpha}{F}_{\,\beta\gamma]}=0\,,
\end{equation}
where $J_{\alpha}$ denotes the electric four current vector.

It is frequently advantageous to represent the electromagnetic field by a vector potential $A_\alpha$ in terms of which the  Maxwell tensor is given as
$F_{\alpha\beta}= \partial_\alpha A_{\beta}-\partial_{\beta} A_{\alpha}$, while the field equations read as 
\begin{equation}\label{m2}
\partial^\alpha\left(\partial_\alpha A_{\beta}-\partial_{\beta} A_{\alpha}\right) =-4\pi\,J_{\beta}\,.
\end{equation}
The choice of $A_\alpha$ is known to be non-unique and two vector potentials $A_\alpha$ and $A'_\alpha$ are physically equivalent if there exists a real function $\chi$ such that 
\begin{equation}\label{vp}
A'_\alpha=A_\alpha + \partial_\alpha \chi\,.
\end{equation}

This freedom is useful in choosing a vector potential suiting to the investigated problem. Start by splitting the vector potential $A_\alpha$ with respect to an inertial reference system\,\footnote{The minus sign in front of the scalar potential is of historical origin. The vector potential $A_\alpha$ itself entered into the discussions much later than the scalar potential.}$^{,}$\footnote{The speed of light will be retained in the equations, and the Gaussian system of units (for its determination see, e.g.\,\cite{jackson}) will be applied, with $\varepsilon_0=1$ and $\mu_0=1$ throughout.}, with coordinates $(t,\vec{r})$ and derivatives $\partial_\alpha=\left(\frac1c\,\partial_t,\vec{\nabla}\right)$, as  $A_\alpha=(-\Phi,\vec{A})$. Then, if the vector potential satisfies the Lorenz gauge condition,\,\footnote{ The index raising and lowering is always done by either of the fixed background metrics $\eta_{\alpha\beta}$ or $\delta_{ij}$ of the Minkowski spacetime or the Euclidean space, 
respectively. Moreover, Einstein's summation convention is used only for identical upper and lower indices.} i.e. \,$\partial^\alpha A_\alpha^L=\frac1c\,\partial_t\Phi_L + \vec{\nabla}\cdot\vec{A}_L=0$ holds for $A_\alpha^L=(-\Phi_L,\vec{A}_L)$, the Maxwell equations (\ref{m2}) simplify to
\begin{equation}
\left(-\frac1{c^2}\,\partial_t^2+\nabla^2\right)\,A_{\alpha}^L= -4\pi\,J_{\alpha} \,,\label{feemL}
\end{equation}
where $\nabla^2$ stands for the Laplace-Beltrami operator.

The other frequently used gauge is the Coulomb gauge, with vector potential $A_\alpha^C=(-\Phi_C,\vec{A}_C)$. It requires the vanishing of the spatial divergence $\vec{\nabla}\cdot\vec{A}_C$, and then the field equations read as 
\begin{eqnarray}
&&\hskip1.73cm \nabla^2\,\Phi_C=-4\pi\, \rho\,,\label{feemCs}\\
&&\left(-\frac1{c^2}\,\partial_t^2+\nabla^2\right)\,\vec{A}_C =-\frac{4\pi}{c}\,\left[\vec{J}
-\frac{1}{4\pi} \vec{\nabla}\left(\partial_t\Phi_C\right)\right]\,,\label{feemCv}
\end{eqnarray}
where $-\rho$ and $\vec{J}$ denote the time and spatial part of the (locally determined) electric four current vector 
$J_\alpha$, respectively. 

Notice that the transverse electric current 
\begin{equation}\label{tc}
\vec{J}_T=\vec{J}-\frac{1}{4\pi} \vec{\nabla}\left(\partial_t\Phi_C\right)\,%,
\end{equation}
extends over all space regardless whether the spatial part $\vec{J}$ of $J_{\alpha}$ is localized or not \cite{jackson}. This is a direct consequence of the fact that the scalar potential $\Phi_C$ is non-local as it is subject to the Poisson equation (\ref{feemCs}) which means that a change in the charge distribution, even if it happens at an astrophysical distance, leads to an instantaneous change in $\Phi_C$. 

\subsection{Potentials of moving point charges}

For a point-like source with electric charge $q$ on an orbit\,\footnote{The tangent vector of this orbit can be given as $\vec{v}(t)=\dot{\vec{R}}(t)$ whereas the pertinent spatial electric current vector reads as $\vec{J}=q \,\vec{v}(t)\,\delta^{(3)}\left[\vec{r}-\vec{R}(t)\right]$.} $\vec{R}=\vec{R}(t)$ the Lorenz gauge vector potential can be given in the familiar Li\'enard-Wiechert form as 
\begin{align}
\Phi_L(t,\vec{r})={}& \,\frac{q}{|\vec{r}-\vec{R}(t_{ret})|}\,\frac1{1-\vec{\beta}(t_{ret})\,\vec{n}(t_{ret})}
\label{wundt29a} \\ %\,.
\vec{A}_L(t,\vec{r})={}&  \frac{q}{|\vec{r}-\vec{R}(t_{ret})|}\,\frac{\vec{\beta}(t_{ret})}{1-\vec{\beta}(t_{ret})\,\vec{n}(t_{ret})}\,, \label{wundt29b}
\end{align}
where the relation between the retarded time $t_{ret}$ and $t$ is
\begin{equation}\label{tret}
F(t,t_{ret})=t-t_{ret}-\frac{|\vec{r}-\vec{R}(t_{ret})|}{c}=0\,,
\end{equation}
$\vec{\beta}(t_{ret})$ is defined as 
\begin{equation}\label{enn}
\vec{\beta}(t_{ret})=\frac{\dot{\vec{R}}(t_{ret})}{c} 
\end{equation} 
and the unit vector
\begin{equation}\label{enn}
\vec{n}(t_{ret})=\frac{\vec{r}-{\vec{R}}(t_{ret})}{|\vec{r}-\vec{R}(t_{ret})|}
\end{equation} 
points from the location of the source $\vec{R}(t_{ret})$ at $t_{ret}$ to the observation point $\vec{r}$. 

\medskip

In case of the Coulomb gauge, for a point charge (\ref{feemCs}) implies that 
\begin{equation}
\Phi_C(t,\vec{r})=\frac{q}{|\vec{r}-\vec{R}(t)|} \label{wundt32} \,,
\end{equation}
and as shown\,\footnote{In order to pass from the units applied in \cite{wundt} to the Gaussian system of units used in our paper $4\pi\epsilon_0$ has to be replaced by $\frac1c$ in (44b) of \cite{wundt}.} in \cite{wundt}
\begin{equation}
\vec{A}_C(t,\vec{r})=\vec{A}_L(t,\vec{r})-q\,c\, \vec{\nabla}\cdot \left[ \int_{t_{ret}}^t  \frac{1}{|\vec{r}-\vec{R}(t')|}\,dt'\right] \,,\label{wundt44b1}
\end{equation}
(see Eq.(44b) of \cite{wundt}). 
 
Note that the second term on the right hand side of (\ref{wundt44b1}) is again an ``action at a distance'' type expression. 

Taking into account (\ref{wundt29b}), along with the relation (\ref{GG}) derived in the Appendix, $\vec{A}_C(t,\vec{r})$ can also be given as 
\begin{equation}\label{gauss_c}
\vec{A}_C(t,\vec{r})=\frac{q}{|\vec{r}-\vec{R}(t_{ret})|}\,\frac{\vec{\beta}(t_{ret})-\vec{n}(t_{ret})}{1-\vec{\beta}(t_{ret})\,\vec{n}(t_{ret})} + q\:c \int_{t_{ret}}^t \frac{\vec{r}-{\vec{R}}(t')}{|\vec{r}-\vec{R}(t')|^3} \,dt'\,.
\end{equation}

As it is discussed in the introduction---in the case of a time independent vector field---the discrepancy arising by the replacement of the proper projection operator $P$ by the simplified transverse one ${\not\hskip-0.1cm P}$ can be given by (\ref{projD}). In the more general case of a moving point charge, as it is verified by (\ref{GG2}) of the Appendix, in classical electrodynamics the error yielded by such a replacement is
\begin{equation}\label{GG3}
\Delta[\vec{A}]= -\frac{q\,\vec{n}(t_{ret})}{|\vec{r}-{\vec{R}}(t_{ret})|} + q\:c \int_{t_{ret}}^t \frac{\vec{r}-{\vec{R}}(t')}{|\vec{r}-\vec{R}(t')|^3} \,dt'\,.
\end{equation} 

\section{Special cases}\label{cases}
\setcounter{equation}{0}

In the succeeding subsections the large distance behavior of $\vec{A}_C$ and $\Delta[\vec{A}]$ will be determined for various motions of a point charge. 

\medskip

%\subsection{The asymptotic form of the generic integral expressions}

However, before proceeding it is rewarding to have a glance at some of the technical difficulties related to the evaluation of the integral term in (\ref{gauss_c}) and (\ref{GG3}). As it will be clear soon even in case of simple one-dimensional motions of a point charge the pertinent integrands may be transcendent functions the integrals of which cannot be given in closed form. Nevertheless, if considerations are restricted to the radiation zone, a careful limiting process allows to determine at least the leading order terms accurately.

\medskip

It is remarkable that whenever the motion of the point particle is restricted to a straight line and the field is also evaluated along this line the unsatisfactory behavior of the simplified transverse projection ${\not\hskip-0.1cm P}$ immediately follows. To see this assume that the above mentioned line coincides with the $x$-axis. Then, in virtue of (\ref{wundt29b}), regardless of the specific form of $R_x=R_x(t)$ the simplified transverse projection ${\not\hskip-0.1cm P}$ yields identically zero result whereas the Coulomb gauge vector potential $\vec{A}_C$ is non-zero unless the point charge is at rest. 

\medskip

Below the constant velocity motion, the oscillatory motion, and the superposition of these two motions of a point charge will be considered. 
By allowing a generic location of the observation point an analogous failure of the simplified transversal projection will be seen to occur whenever the orbit of the point charge is unbounded with respect to the observer's reference frame. 

\subsection{Motion with constant velocity}\label{cases1}
\setcounter{equation}{0}

Start with the simplest possible motion of constant velocity. Accordingly, we shall assume\,\footnote{The vanishing of $\vec{R}$ at $t=0$ makes use of the freedom we have in choosing the origin of the reference system. Note also that this freedom is not specific to the particular case considered in this subsection so it will be applied in the other two cases, as well.} that the point charge moves along the $x$-axis with speed $\vec{v}=(\varepsilon v,0,0)$, i.e.
\begin{equation} 
\vec{R}=\vec{R}(t)=\vec{v}\,t = \left\{ \begin{array} {l} R_x=\varepsilon\,v \,t  \,;\\ R_{\{y,z\}}=0\,, \end{array}
\right.\label{Rx1}
\end{equation} 
where $\varepsilon$ is a sign taking the values $+1$ for forward and $-1$ for backward motion, respectively. Note also that hereafter the notation ${\{y,z\}}$ is used to indicate results relevant for the $y$ and $z$ components.   

\medskip

Then, by using the notation $\vec{r}=r\,\vec{e}$, where $\vec{e}$ denotes the unit spatial vector pointing from the origin to the location of the observation point, the expressions $\vec{r}-{\vec{R}}(t')$ and $|\vec{r}-{\vec{R}}(t')|$ can 
be given as 
\begin{equation}
\vec{r}-{\vec{R}}(t')=r\,\vec{e} - \vec{v}\,t'\,,\ \ {\rm and}\ \ \ |\vec{r}-{\vec{R}}(t')|= \sqrt{(r\,e_x - \varepsilon\,v\,t')^2 + r^2\,(1 - e_x^2)}\,,
\end{equation}
where the relation $e_x^2+e_y^2+e_z^2=1$ has been applied, while for the integrals
\begin{align}
{}& \int_{t_{ret}}^t \frac{(\vec{e}-{\vec{R}}(t'))_x}{|\vec{r}-\vec{R}(t')|^3} \,dt' =\frac{1}{\varepsilon\,v}\left[\frac{1}{\sqrt{r^2 + t^2\, v^2 - 2 \,r \,\varepsilon \,e_x \,v\,t}}-\frac{1}{\sqrt{r^2 + t_{ret}^2\, v^2 - 2\, r\,\varepsilon\, e_x\, v\, t_{ret}}}\right]\,,\label{int_unix} \\ 
{}& \int_{t_{ret}}^t \frac{(\vec{e}-{\vec{R}}(t'))_{\{y,z\}}}{|\vec{r}-\vec{R}(t')|^3} \,dt'\nonumber \\ {}& \hskip1.5cm = -\frac{e_{\{y,z\}}}{r\,\varepsilon\,v\,(1-e_x^2)}\left[\frac{r\,e_x - \varepsilon\,v\,t}{\sqrt{r^2 + t^2\, v^2 - 2 \,r \,\varepsilon \,e_x \,v\,t}}-\frac{r\,e_x - \varepsilon\,v\,t_{ret}}{\sqrt{r^2 + t_{ret}^2\, v^2 - 2\, r\,\varepsilon\, e_x\, v\, t_{ret}}}\right]\label{int_uniyz}
\end{align}
can be seen to hold. Taking into account (\ref{tret}) the time of observation $t$ can be expressed as 
\begin{equation}
t = t_{ret} + \frac{\sqrt{(r\,e_x - \varepsilon\,v\,t_{ret})^2 + r^2\,(1 - e_x^2)}}{c}\,
\end{equation}
and determining the leading $\frac1r$-terms of the components of $\vec{A}_C$, relevant for a fixed value $t_{ret}$, we get 
\begin{align}
A_{C}^x = &{}-\frac{q}{{r\,\varepsilon\, \beta}}\left[\frac{1-\beta^2}{1 - \varepsilon\,\beta\, e_x} - 
\frac{1}{\sqrt{1 + \beta^2 - 2\, \beta\, e_x }}\right]+\mycal{O}(r^{-2})\,,\label{ACx} \\
A_C^{\{y,z\}} = &{} \frac{q}{{r\,\varepsilon\, \beta}}\hskip-.1cm\cdot\hskip-.1cm\frac{ e_{\{y,z\}}\,\left(e_x-\varepsilon\,\beta\right)}{1-e_x^2}
\left[\frac{1}{1-\beta\,\varepsilon\,e_x} - \frac{1}{\sqrt{1 + \beta^2 - 2 \,\varepsilon\,\beta\, e_x }}\right]+\mycal{O}(r^{-2})\,,\label{ACz}
\end{align}
where $\beta=\frac vc$. Note that the specific value of $t_{ret}$ appears only in the higher order terms.

\medskip

By evaluating the generic expression (\ref{GG3}) for  $\Delta[\vec{A}]$, using again the same fixed value $t_{ret}$, we get
\begin{align}
\Delta[\vec{A}]^x = {}& -\frac{q}{{r\,\varepsilon\, \beta}} \left[ 1+ \varepsilon\,\beta\, e_x - \frac{1}{\sqrt{
   1 + \beta^2 - 2 \,\varepsilon\,\beta\, e_x }}\right]+\mycal{O}(r^{-2})\,,\label{DeltaAx}\\
\Delta[\vec{A}]^{\{y,z\}} = {}& \frac{q}{{r\,\varepsilon\, \beta}}\hskip-.1cm\cdot\hskip-.1cm\frac{ e_{\{y,z\}}}{1-e_x^2}
\left[e_x -\beta\,\varepsilon\left(1 - e_x^2\right) + \frac{\varepsilon\,\beta-e_x}{\sqrt{
   1 + \beta^2 - 2 \,\varepsilon\,\beta\, e_x }}\right]+\mycal{O}(r^{-2})\,.\label{DeltaAz}
\end{align}

By inspecting the above relations it gets immediately transparent that $\Delta[\vec{A}]$ is of the same order as  $\vec{A}_C=P[\vec{A}_L]$. Nevertheless, one could argue that this example is not representative as there is no electromagnetic radiation associated with the uniform motion of a point charge. To this end it is important to be mentioned that---as it will be demonstrated in subsection \ref{cases3}---an analogous discrepancy of the proper and the simplified transversal projections occurs when a harmonic oscillation is superimposed on the uniform motion and then radiation is also involved. 

\subsection{Oscillatory motion}\label{cases2}

Consider now a harmonic oscillation of a point charge. Accordingly, it will be assumed that
\begin{equation} 
\vec{R}=\vec{R}(t)= \left\{ \begin{array} {l} R_x=a\,\sin( \omega\,t)  \,;\\ R_{\{y,z\}}=0\,, \end{array}
\right.\label{Rx2}
\end{equation}
i.e.\,the charge oscillates around the origin of the reference frame of the observer along the $x$-axis with amplitude $a$. 

\medskip

With (\ref{Rx2}) the integral term appearing in (\ref{gauss_c}) and (\ref{GG3}) becomes
\begin{equation}\label{int_osc0}
\int_{t_{ret}}^t \frac{{r_x}-{{R_x}}(t')}{|\vec{r}-\vec{R}(t')|^3} \,dt'=\frac{1}{r^2}\int_{t_{ret}}^t \frac{{e_x} - \frac ar \,\sin( \omega\,t') }{\left[1 -  \frac {2 \,\,e_x a}r \,\sin( \omega\,t') + \frac {a^2}{r^2} \,\sin^2( \omega\,t')\right]^{\frac32}}\,dt'\,.
\end{equation}

Our aim as before is to determine the asymptotic behavior of this integral along a null line characterized by a specific value of $t_{ret}$.  
In doing so notice first that for any fixed value of $r$ the integrand on the right hand side 
\begin{equation}\label{int_osc2}
{\mycal F}(t',r)=\frac{{e_x} - \frac ar \,\sin( \omega\,t') }{\left[1 -  \frac {2 \,\,e_x a}r \,\sin( \omega\,t') + \frac {a^2}{r^2} \,\sin^2( \omega\,t')\right]^{\frac32}}
\end{equation}
is a bounded periodic function of $t'$.

Writing now $t -t_{ret}$ as 
\begin{equation}\label{tret2}
t -t_{ret}= N\cdot T+\Delta t\,, 
\end{equation}
where $N$ is a sufficiently large integer and $\Delta t$ is smaller than $T$, the period of the oscillation, the integral  $\int_{t_{ret}}^t {\mycal F}(t',r) \,dt'$ can be given as 
\begin{equation}\label{int_osc}
\int_{t_{ret}}^t {\mycal F}(t',r) \,dt'=\sum_{i=1}^{N}\int_{t_{ret}+(i-1)\cdot T}^{t_{ret}+ i\cdot T} {\mycal F}(t',r) \,dt'+ \int_{t_{ret}+N\cdot T}^{t_{ret}+ N\cdot T + \Delta t} {\mycal F}(t',r) \,dt' \,.  
\end{equation}

Due to the periodicity of the integrand in $t'$ the integrals on the right hand side of (\ref{int_osc}) are defined with respect to fixed and finite intervals. Thereby, we may apply Theorem 9.42 of \cite{rudin} ensuring that whenever the integral $\int_{a}^b {\mycal F}(t',r) \,dt'$  exists for the closed interval $[a,b]$ in $\mathbb{R}$, and the integrand ${\mycal F}(t',r=\rho^{-1})$ is (at least) a $C^1$ function of $\rho$, the relation 
\begin{equation}\label{rudin}
\partial_\rho\left[\int_{a}^b {\mycal F}(t',\rho^{-1}) \,dt'\right]=\int_{a}^b \partial_\rho\left[{\mycal F}(t',\rho^{-1})\right] \,dt'\,,
\end{equation}
holds. This, in particular, implies that as far as we are only interested in the asymptotic behavior of these terms in the $r\rightarrow \infty$ limit and whenever ${\mycal F}(t',r)$ is sufficiently smooth the $\frac1{r}$-series expansion of 
\begin{equation}\label{rudin}
\int_{a}^b {\mycal F}(t',r) \,dt'
\end{equation}
is equal to that of the integral of the $\frac1{r}$-series expansion of ${\mycal F}(t',r)$. 
By applying these observations and taking into account that the $\frac1{r}$-series expansion of ${\mycal F}(t',r)$ in (\ref{int_osc2}) reads as 
\begin{equation}\label{int_osc2_ser}
%Series\left[{\mycal F}(t',r);r,3\right]=
e_x-\frac ar\, \sin\left(\omega\, t'\right)\, \left[1-3\,e_x^2\right]-\frac {3\,a^2}{2\,r^2}\, \sin^2\left(\omega\, t'\right)\, \left[3-5\,e_x^2\right] + \mycal{O}(r^{-3})
\end{equation}
the integral on the right hand side of (\ref{int_osc0}) can be evaluated. 

\medskip

By combining this with (\ref{gauss_c}), and taking into account $N\approx \frac{t -t_{ret}}{T}\approx \frac{r}{c\,T}$, it is straightforward to verify that 
\begin{eqnarray}
A_{C}^x &\hskip-.3cm=&\hskip-.3cm -\frac {q\,a\,\omega\,\left[1-e_x^2\right]\,\cos(\omega\,t_{ret})}{r\left[c+a\,\omega\,\cos(\omega\,t_{ret})\,e_x\right]}%+ q\,a\,\sin(\omega\,t_{ret})\times \nonumber \\ 
%&&\hskip-3.3cm
%\frac{-2 c^2 + 2\, a^2 \,\omega^2\, \cos^2(\omega\,t_{ret}) + 2\,a\,c\,\omega\,\cos(\omega\,t_{ret})\,e_x + 6\, c^2\,e_x^2 + 6\, a\, c\,\omega\, \cos(2\,\omega\,t_{ret})\,e_x^3 + 2\,a^2\,\omega^2\, \cos(2\,\omega\,t_{ret})^2 \,e_x^4
%}{2\,r^2\left[c+a\,\omega\,\cos(\omega\,t_{ret})\,e_x\right]^2}
 + \mycal{O}(r^{-2})\,,\label{ACxosc} \\
A_{C}^{\{y,z\}} &\hskip-.3cm=&\hskip-.3cm \frac {q\,a\,\omega\,e_x\,e_{\{y,z\}}\,\cos(\omega\,t_{ret})}{r\left[c+a\,\omega\,\cos(\omega\,t_{ret})\,e_x\right]}
 + \mycal{O}(r^{-2})\,.\label{ACyzosc}
\end{eqnarray}

\medskip

Similarly, in virtue of (\ref{GG3}), the leading order expressions for the error term $\Delta[\vec{A}]$ can be seen to take the form 
\begin{eqnarray}
\hskip-.5cm\Delta[\vec{A}]^x&\hskip-.3cm=&\hskip-.3cm-\frac{2\,q\,a\,e_x^2\sin(\omega\,t_{ret})}{r^2}
%-\frac{q\,a\,e_x^2}{c\,r^2}\,\sin(\omega\,t_{ret})\nonumber \\ 
%&&\hskip-.3cm+ \frac{q\,a\,e_x^2\left(1- 3\,e_x^2\right)}{2\,c\,\omega\,r^3}\left[a\,\omega\,e_x\,\sin^2(\omega\,t_{ret})\pm4\,c\,\pi \right] 
+\mycal{O}(r^{-3})\,,\label{DeltaAxosc}\\
\hskip-.5cm\Delta[\vec{A}]^{\{y,z\}}&\hskip-.3cm=&\hskip-.3cm-\frac{2\,q\,a\,e_x\,e_{\{y,z\}}\sin(\omega\,t_{ret})}{r^2}\,%\nonumber \\ 
%&&\hskip-.3cm+ \frac{q\,a\,e_{\{y,z\}}}{2\,c\,\omega\,r^3}\left[a\,\omega\left(1- 3\,e_x^2\right)\,\sin^2(\omega\,t_{ret})\pm12\,c\,\pi\,e_x\right] 
+\mycal{O}(r^{-3})\,.\label{DeltaAzosc}
\end{eqnarray}

The last two relations verify that for the case of an oscillatory motion of a point charge with a fixed center of oscillation with respect to the observer's reference system the error $\Delta[\vec{A}]$ is of higher order than the Coulomb gauge vector potential itself---the latter falls off as $\frac1{r}$---hence the error caused by the replacement of the proper projection by the simplified transversal one is negligible in the asymptotic region. 

\subsection{Combination of uniform and oscillatory motions}\label{cases3}

In this subsection it will be shown that the discrepancy between the proper and simplified transversal projections also occurs when radiation is present, in particular when a harmonic oscillation is superimposed on the uniform motion. It is noteworthy that this superposition can also be looked upon as an oscillation around a center moving with constant speed. The relevance of this model is suggested by the dynamical character of our Universe.

\medskip

Accordingly, we shall assume that the center of oscillation moves with constant velocity $v$ with respect to the reference system of the observer%, i.e.~ 
\begin{equation} 
\vec{R}=\vec{R}(t)= \left\{ 
\begin{array} {l} R_x=\varepsilon\, v\,t+a\,\sin( \omega\,t)  \,;\\ R_{\{y,z\}}=0\,, 
\end{array}
\right.\label{Rx3}
\end{equation}
with $\varepsilon$ defined as in (\ref{Rx1}).
%i.e.\,the charge oscillates with amplitude $a$ around a center moving with constant speed $\varepsilon\,\vec{v}$ such that both of these motions occur along the $x$-axis, while all the other components of $\vec{R}=\vec{R}(t)$ vanish. 

\medskip

A glance at the right hand side of 
\begin{equation}\label{int_osc02}
\int_{t_{ret}}^t \frac{{r_x}-{{R_x}}(t')}{|\vec{r}-\vec{R}(t')|^3} \,dt'=\frac{1}{r^2}\int_{t_{ret}}^t \frac{{e_x} - \frac {\varepsilon\, v\,t'+ a \,\sin( \omega\,t') }r}{\left[1 -  \frac {2 \,\,e_x  \left(\varepsilon\, v\,t'+ a \,\sin( \omega\,t') \right)}r + \frac {\left(\varepsilon\, v\,t'+ a \,\sin( \omega\,t') \right)^2}{r^2}\right]^{\frac32}}\,dt'\,.
\end{equation}
makes immediately clear that the determination of the rate of the fall off of this integral is complicated. In this respect it turned out to be rewarding to consider first the integral of the difference 
\begin{equation}\label{int_osc03}
\frac{{e_x} - \frac {\varepsilon\, v\,t+ a \:\sin( \omega\,t) }r}{\left[1 -  \frac {2 \,\,e_x  \left(\varepsilon\, v\,t+ a \:\sin( \omega\,t) \right)}r + \frac {\left(\varepsilon\, v\,t+ a \,\sin( \omega\,t) \right)^2}{r^2}\right]^{\frac32}}- \frac{{e_x} - \frac {\varepsilon\, v\,t }r}{\left[1 -  \frac {2 \,\,e_x  \,\varepsilon\, v\,t}r + \frac { v^2\,t^2}{r^2}\right]^{\frac32}}\,.
\end{equation}
Notice that the subtracted term is the integrand applied in case of the pure uniform motion. What makes the use of this difference really advantageous is that the amplitude of the oscillation of this difference (for a fixed value of $r$) falls off at least as fast as $\frac1{t^3}$.\,\footnote{To see this a Taylor expansion in powers of $\frac1t$ has been applied keeping $t$ in $\sin(\omega t)$ fixed in $\varepsilon\, v\,t+ a \:\sin( \omega\,t)$ since $|a \:\sin( \omega\,t)|\ll |\varepsilon\, v\,t|$ while $\frac1t\rightarrow 0$. The validity of the  $\frac1{t^3}$ approximation has also been verified by numerical evaluations.} This, along with the fact that we have the factor $\frac{1}{r^2}$ in front of the integral on the right hand side of (\ref{int_osc02}) and $t-t_{ret}$ goes as $\frac{r}{c}$ verifies that unless one is interested in the second or higher order contributions in $\frac{1}{r}$ the integral of this difference in (\ref{int_osc03}) may be neglected. Thereby, the leading order of the fall off 
rate of the integral 
\begin{equation}\label{int_osc04}
\int_{t_{ret}}^t \frac{{r_x}-{{R_x}}(t')}{|\vec{r}-\vec{R}(t')|^3} \,dt'=\frac{1}{r^2}\int_{t_{ret}}^t\frac{{e_x} - \frac {\varepsilon\, v\,t }r}{\left[1 -  \frac {2 \,\,e_x  \,\varepsilon\, v\,t}r + \frac { v^2\,t^2}{r^2}\right]^{\frac32}} + \mycal{O}(r^{-2})
\end{equation}
is exactly the same as that of (\ref{int_unix}). 

\medskip

In determining the fall off rates of $\vec{A}_C$ and $\Delta[\vec{A}]$ one also has to evaluate the first terms of the right hand sides of (\ref{gauss_c}) and (\ref{GG3}). By combining these terms with (\ref{int_osc04}), along with the leading order terms of (\ref{ACx}) and (\ref{ACz})---which will be refereed to as $A_{C, ({\tiny\ref{ACx}})}^x$ and $A_{C, ({\tiny\ref{ACz}})}^{\{y,z\}}$---the leading order terms of $\vec{A}_C$ read as 
\begin{align}
A_{C}^x = &{} A_{C}^x{}_{({\tiny\ref{ACx}})} + \frac{q\,a\,\omega\,[1- e_x^2]\, \cos(\omega\,t_{ret})}{r\,(1-\beta\,\varepsilon\,e_x)\,[c-c\,\beta\,\varepsilon\,e_x - a\,\omega\,e_x \cos(\omega\,t_{ret})]}+\mycal{O}(r^{-2}) \,,\label{ACx2} \\
A_C^{\{y,z\}} = &{} A_{C, ({\tiny\ref{ACz}})}^{\{y,z\}} -\frac{q\,a\,\omega\, e_x\, e_{\{y,z\}}\, \cos(\omega\,t_{ret})}{r\,(1-\beta\,\varepsilon\,e_x)\,[c-c\,\beta\,\varepsilon\,e_x - a\,\omega\,e_x \cos(\omega\,t_{ret})]}+\mycal{O}(r^{-2})\,,\label{ACz2}
\end{align}
whereas the leading order terms of the components of $\Delta[\vec{A}]$ are exactly the same as those in (\ref{DeltaAx}) and (\ref{DeltaAz}). 

The above relations implies that in the current case, where radiation is also involved, even in the asymptotic region the discrepancy $\Delta[\vec{A}]$ is of the same order as $\vec{A}_C=P[\vec{A}_L]$. 

\section{Final remarks}\label{con}
\setcounter{equation}{0}

In electrodynamics one of the most frequently used gauges is the Coulomb gauge. The transformation to this gauge from any other gauge is an important chapter of the lecture courses on electrodynamics. The present note contains a warning about the use of the simplified projection operator (\ref{projN}) when carrying out the transformation.

\medskip

In this paper some properties of the proper and simplified transverse projection operators in electrodynamics were studied. As discussed in the introduction it is widely held that the discrepancy of the action of the proper and of the simplified transversal projections is asymptotically negligible. More precisely, it is usually claimed that although the simplified transverse projection may have some error in the near or intermediate zone, this error should be negligible in the radiation zone (see, e.g.~the paragraph below Eq.\,(3.16) in \cite{jackson2}). In investigating the validity of these expectations the particular cases of the constant velocity motion, the oscillatory motion, and the superposition of these two motions of a point charge were studied in some details. 

\medskip

In the case of purely oscillatory motion---i.e.~when the center of oscillation of the point charge is fixed with respect to the observer's reference frame---the generic expectation was verified by our analysis in subsection \ref{cases2}. 

\medskip

Note, however, that when the interminable motion of the source is spatially unbounded with respect to the observer's reference frame the situation is different. According to our main result in this case the replacement of the proper projection operator by the simplified transverse one yields, even in the radiation zone, an erroneous result with error of the same order as the proper Coulomb gauge vector potential itself.

\medskip

Having these results it may be reasonable to ask what may be responsible for the failure of our intuition when we apply the simplified transversal projection instead of the proper one. In answering this question it is worth recalling that the simplified projection refers merely to local fields such as the vector potential in Lorenz gauge (\ref{wundt29b}) and the normal vector (\ref{enn}) pointing from the source to the observer. As opposed to this both the Coulomb gauge vector potential and the proper projection operator involve an integral term---see the second terms in (\ref{gauss_c}) and (\ref{GG3})---which is of an action at a distance type. This integral is asymptotically non-negligible when the interminable relative motion of the source and the observer is unbounded. Therefore not the reported discrepancies but their irrelevance in case of bounded motions is remarkable.

\medskip

As an interesting possible implication of the above observations recall that an analogous replacement of the proper projection operator by the simplified transverse one is applied in gravitational wave (GW) physics in determining the metric perturbation in ``transverse traceless'' ($TT$) gauge (see, e.g.~\cite{racz} for more details). Based on the dynamical character of our Universe it is highly probable that the gravitational wave sources are not fixed with respect to our GW detectors.  This, in virtue of our results in subsection \ref{cases3}, may necessitate a careful revision of some of the arguments based on the use of the analog of the simplified transverse projection in linearized theory of gravity.

\medskip

Finally, it is worth emphasizing that in electrodynamics all the troubles yielded by the replacement of the proper projection operator by simplified transversal one goes away once gauge independent quantities are applied. Analogously, in the linearized theory of gravity one could avoid all the technical difficulties in describing the propagation of gravitational waves and their effects on GW detectors by using the curvature tensor instead of the gauge dependent metric perturbations (see the discussions in \cite{racz, david}).

\section*{Acknowledgments}

%This research was supported in part by the Die Aktion \"Osterreich-Ungarn, Wissenschafts- und Erziehungskooperation grant 87\"ou16. 
One of us, IR was supported by the European Union and the State of Hungary, co-financed by the European Social Fund in the framework of T\'AMOP-4.2.4.A/2-11/1-2012-0001 ``National Excellence Program''.

\appendix
\section{Appendix}%\label{Appendix A}
\renewcommand{\theequation}{A.\arabic{equation}}
\setcounter{equation}{0}

This Appendix is to provide a justification of (\ref{GG3}). 

Note first that by making use of (\ref{tret}), along with the implicit function theorem, the relations 
\begin{equation}\label{dtretdt}
\frac{\partial t_{ret}}{\partial t}=-\left(\frac{\partial F(t,t_{ret})}{\partial t_{ret} }\right)^{-1}\cdot \frac{\partial F(t,t_{ret})}{\partial t }=  \left(1-\frac{\dot{\vec{R}}(t_{ret})}{c}\, \frac{\vec{r}-{\vec{R}}(t_{ret})}{|\vec{r}-\vec{R}(t_{ret})|}\right)^{-1}\,,
\end{equation}
and 
\begin{equation}\label{gradtret3}
\vec{\nabla} t_{ret} =-\frac{\vec{r}-{\vec{R}}(t_{ret})}{c\,|\vec{r}-{\vec{R}}(t_{ret})|-\left[\left(\vec{r}-{\vec{R}}(t_{ret})\right)\cdot\dot{\vec{R}}(t_{ret}) \right]}%\,.
\end{equation}
can be seen to hold.\,\footnote{Notice that (\ref{gradtret3}) is equivalent to (A.5) of \cite{wundt}.}

Consider now the second term on the right hand side of (\ref{wundt44b1}). In doing so introduce the notation
\begin{equation}\label{G}
\vec{\nabla} \left[G(t,t_{ret},\vec{r}) \right] =  \vec{\nabla} \left[ \int_{t_{ret}}^t  \frac{1}{|\vec{r}-\vec{R}(t')|}\,dt'\right] \,,
\end{equation}
where the retarded time is supposed to be given by a function $t_{ret}=t_{ret}(t,\vec{r})$.
Then the right hand side of (\ref{G}) can also be written as 
\begin{equation}\label{G2}
\partial_\alpha \left[G(t,t_{ret}(t,\vec{r}),\vec{r}) \right] =  \left(\frac{\partial G}{\partial t_{ret}} \right) \left(\partial_\alpha t_{ret}\right)+``\,\partial_\alpha G\," \,,
\end{equation}
where 
\begin{equation}\label{G04}
``\,\partial_\alpha G\,"=\int_{t_{ret}}^t \partial_\alpha  \left[\frac{1}{|\vec{r}-\vec{R}(t')|}\right]\,dt'\,.
\end{equation}

Making use of the properties of integrals as functions of boundary values we get 
\begin{equation}\label{G3}
\frac{\partial G}{\partial t_{ret}}=-\frac{1}{|\vec{r}-{\vec{R}}(t_{ret})|}\,
\end{equation}
whereas (\ref{G04}) reads as
\begin{equation}\label{G4}
``\,\partial_\alpha G\," = -\int_{t_{ret}}^t \frac{(x^\beta-X^\beta(t'))\,{\delta_\alpha}^\beta }{|\vec{r}-\vec{R}(t')|^3}\,dt'= -\int_{t_{ret}}^t \frac{\vec{r}-{\vec{R}}(t')}{|\vec{r}-\vec{R}(t')|^3}\,dt' \,.
\end{equation} 

In virtue of (\ref{gradtret3}) and (\ref{G2}) - (\ref{G4}) we also have that 
\begin{eqnarray}\label{GG}
&&\vec{\nabla} \left[ \int_{t_{ret}}^t  \frac{1}{|\vec{r}-\vec{R}(t')|}\,dt'\right] = \frac1c\,\frac{\vec{r}-{\vec{R}}(t_{ret})}{|\vec{r}-{\vec{R}}(t_{ret})|^2} 
\left(1-\frac{\dot{\vec{R}}(t_{ret})}{c}\, \frac{\vec{r}-{\vec{R}}(t_{ret})}{|\vec{r}-\vec{R}(t_{ret})|}\right)^{-1} \nonumber \\ 
&& \phantom{\vec{\nabla} \left[ \int_{t_{ret}}^t  \frac{1}{|\vec{r}-\vec{R}(t_{ret})|}\,dt'\right] =}- \int_{t_{ret}}^t \frac{\vec{r}-{\vec{R}}(t')}{|\vec{r}-\vec{R}(t')|^3}\,dt'\,.
\end{eqnarray}

Then taking into account (\ref{projD}) and assuming that $\vec{A_L}$ is known we get 
\begin{align}\label{GG2}
\Delta[\vec{A}]= {}& (\vec{n}(t_{ret})\cdot\vec{A_L})\,\vec{n}(t_{ret}) \nonumber \\ {}& -q\,\left[\,\frac{\vec{n}(t_{ret})}{|\vec{r}-{\vec{R}}(t_{ret})|} 
\left(1-\frac{\vec{n}(t_{ret})\cdot\dot{\vec{R}}(t_{ret})}{c}\right)^{-1}- c\, \int_{t_{ret}}^t \frac{\vec{r}-{\vec{R}}(t')}{|\vec{r}-\vec{R}(t')|^3}\,dt' \right]\,,
\end{align} 
where the unit spatial vector $\vec{n}(t_{ret})$ is given in (\ref{enn}). 
Finally, by substituting (\ref{wundt29b}) and using some algebra, it can be seen that (\ref{GG3}) holds.

\end{document}